\documentclass{llncs}
\usepackage{epsfig}
\usepackage[hyphens]{url}
\usepackage{appendix}
\usepackage{listings} 

\title{A Critique of Immunity Passports and W3C Decentralized Identifiers}

\titlerunning{A Critique of Immunity Passports}  
\author{ Harry Halpin}

\authorrunning{Halpin} 

\institute{K.U. Leuven, ESAT/COSIC, \\
 Kasteelpark Arenberg 10, 3001 Leuven, Belgium\\
\email{harry.halpin@esat.kuleuven.be} \\
}

\begin{document}
 
\maketitle              

\begin{abstract}
  Due to the widespread COVID-19 pandemic, there has been a push for `immunity passports' and even technical proposals. Although the debate about the medical and ethical problems of immunity passports has been widespread, there has been less inspection of the technical foundations of immunity passport schemes. These schemes are envisaged to be used for sharing COVID-19 test and vaccination results in general. The most prominent immunity passport schemes have involved a stack of little-known standards, such as Decentralized Identifiers (DIDs) and Verifiable Credentials (VCs) from the World Wide Web Consortium (W3C). Our analysis shows that this group of technical identity standards are based on under-specified and often non-standardized documents that have substantial security and privacy issues, due in part to the questionable use of blockchain technology. One concrete proposal for immunity passports is even susceptible to dictionary attacks. The use of `cryptography theater' in efforts like immunity passports, where cryptography is used to allay the privacy concerns of users, should be discouraged in standardization. Deployment of these W3C standards for `self-sovereign identity' in use-cases like immunity passports could just as well lead to a dangerous form identity totalitarianism.
\end{abstract}

\keywords{immunity passports, Decentralized Identifiers, Verifiable Credentials, W3C, security, privacy, standardization}

\section{Introduction}
\label{sec:intro}

With the outbreak of COVID-19 in 2020, there became a surge of interest in what are called `immunity passports' and various technical proposals to implement these passports in order to allow people to work and travel. In fact, one academic paper claims that in terms of COVID-19 immunity passports, there's ``an app for that''~\cite{eisenstadt}.  Indeed, given the scale of the crisis inflicted on the world by COVID-19, it should not be surprising that the vision of a digital application that could allow people to return to work and travel would be appealing to many governments, and some governments such as Chile\footnote{\url{https://www.thelancet.com/journals/lancet/article/PIIS0140-6736(20)31096-5/fulltext}} and El Salvador are continuing to propose COVID-19 immunity passports.\footnote{\url{https://www.premiumtimesng.com/coronavirus/408007-el-salvador-to-give-immunity-passport-to-those-who-recovered-from-covid-19.html}} A vaguely UN-related organization called `ID2020' has begun to certify digital immunity passports by companies such as BLOCK BioScience as a ``good ID'' to sell to governments.\footnote{\url{https://www.biometricupdate.com/202008/id2020-certifies-blok-bioscience-immunity-passport-with-self-sovereign-approach-to-digital-id}} Therefore, even though there is yet no evidence that a negative test after COVID-19 infection presents long-lived immunity, there appears to be momentum for digital immunity passports. While there has already been considerable medical, ethical, and legal objections to immunity passports~\cite{nature}, there has yet to be a technical analysis of the proposed standards used by immunity passports. 

Although the possible social benefits and harms of immunity passports are outside of the scope of a technical analysis and so will only be briefly discussed, it should go without saying that the status of a person's COVID-19 antibody test results are sensitive personal data. Therefore, a technical analysis should provide a comprehensive overview of the privacy and security properties of any given immunity passport system. The particular use-case of immunity passports and the wider context of digital identity is reviewed in Section~\ref{sec:immunity}. Then each component of the proposed technical architecture of the COVID Credentials Initiative (CCI),\footnote{\url{https://www.covidcreds.com/}} which has already gained considerable media coverage\footnote{\url{https://www.coindesk.com/covid-19-immunity-passport-unites-60-firms-on-self-sovereign-id-project}} and claims over a hundred members, will be inspected. CCI currently has at least fifteen members building on World Wide Web Consortium (W3C) standards, a membership-driven standards bodies known for such standards as XML and early versions of HTML. Note that while we use the term `immunity passport' (as well as `immunity credential,' the digital implementation of an `immunity passport') in this analysis, our usage of the term and analysis also covers antibody test results in general, including vaccination test results. 

For each component, we will first present the standard and then a critique. First, the W3C Verifiable Credentials Data Model is analyzed in Section \ref{sec:vc} and then W3C Decentralized Identifier standards in Section \ref{sec:did}. This lets us analyze in detail the technical architecture put forward when used in an actual user-facing immunity passport app that is built on these W3C standards in Section~\ref{sec:app}. In conclusion, we'll review the dangers of unscoped and premature optimization in standardization and ways to prevent emergencies such as COVID-19 from leading to the abuse of security standardization in Section~\ref{sec:abuse}, before concluding with our vision for next steps and future research in Section~\ref{sec:con}.

\section{Immunity Passports: The Killer Use-case of Digital Identity?}
\label{sec:immunity}

\subsection{Immunity Passports}
An immunity passport can be thought of as a kind of digital \emph{credential} that contains information needed to determine if an individual has contracted a particular disease or not and  whether or not they may be immune to future development of the disease. The concept of immunity credentials were inspired by the idea of a digital update to the well-known `Carte Jeune' paper cards needed to prove yellow fever vaccination and so authorize travel in certain countries. However, it has been noted the development of yellow fever is not analogous to COVID-19, but rather to measles, as measles became widespread in the population and vaccination was mandatory (and so no paper card for travel was needed)~\cite{nature}. However, a possible future COVID-19 vaccine could be proposed as a kind of `immunity passport' to allow travel and work.  

An immunity credential for COVID-19 would have to contain the measurement of antibodies taken by a particular institution at a particular time. The serological antibody measurements include both Immunoglobulin M (the largest antibody) and Immunoglobulin G (the most common antibody) responses to COVID-19~\cite{larremore2020implications}. There are other tests that are commonly available that include tests for the presence of COVID-19 DNA, such as  polymerase chain reaction (PCR) tests and antigen tests, but these tests only detect if COVID-19 is active and so do not detect whether or not a person has been infected or vaccinated in the past. 

\textbf{Critique.} There are a large number of critiques of the very concept of immunity credentials and we will only overview a selection. It has been thought that having COVID-19 antibodies would lead to immunity for a period of time, although recently there have been documented cases of reinfection~\cite{larremore2020implications}. Of course, the primary critique is that the immunoglobulin tests do not actually provide any level of immunity medically and that, even if they did, the level of false positives and negatives is still too  high to be acceptable~\cite{nature}. The effects of antibodies is to confer some level of immunity for an unknown, but likely very short, period of time~\cite{nature}. Medical research is still unclear how long antibodies could prevent COVID-19, if at all, and whether the COVID-19 virus itself is mutating such that antibodies are even relevant~\cite{larremore2020implications}.

Social effects of immunity credentials are possibly dangerous as immunity credential holders could become an `immunity elite' with increased social stratification from those without certificates, violating existing laws on discrimination in many countries~\cite{kaminer2020discrimination}. One dangerous outcome would be people attempting to infect themselves in order to gain the advantages conferred by immunity~\cite{nature}. Although the term `immunity passport' has gotten such a bad reputation that it may seem unlikely to be implemented, the push for COVID-19 vaccination could cause the idea of a digital certificate for COVID-19 test results to be revived in the near future. A digital certificate for COVID-19 test results would be subject to the same critique on a social level as an immunity passports.

\subsection{Digital Identity}

The field of digital identity has long existed in national standards for identity databases, but became a focus of standardization after the tremendous success of the World Wide Web led many people to become interested in the prospects for an internet-enabled digital identity. On the one hand, the Web used an identity system, the domain name system (DNS), which has been wildly successful at providing unique names for web-sites. On the other hand, the internet did not include any provision for providing unique identities for people and organizations that worked across websites, with cookies as a means for implicitly establishing user identities being added later in the development of the Web by browsers.

An augmented social network built on digital identity was theorized as possibly leading to another cycle of innovation and profit as powerful as the original Web~\cite{jordan2003augmented}. Although obviously limited and ontologically problematic~\cite{halpin:standards}, digital identity is usually construed as an unique identifier connected to a set of attributes, such as a name, age, and citizenship. `Self-sovereign' identity gave the identified individual themselves the ability to control these attributes, as opposed to a centralized government or corporation.\footnote{\url{https://www.lifewithalacrity.com/2016/04/the-path-to-self-soverereign-identity.html}}

The goal of digital identity is to avoid identifier collision by assigning globally unique identifiers. The first proposed standard that tried to assign humans and organizations permanent identifiers was eXtensible Resource Identifiers (XRIs) at the OASIS standards body.\footnote{OASIS is the Organization for the Advancement of Structured Information Standards, was mostly known for various XML related standards and allowing, unlike the W3C and IETF, patent licensing fees in standards.} Individuals are given identifiers such as \emph{+david}.\footnote{\url{https://www.oasis-open.org/committees/download.php/15376/xri-syntax-V2.0-cs.html}} Like DNS, XRIs are resolved by XDI (XRI Data Interchange), which would then retrieve an XRDS (Extensible Resource Descriptor Sequence) with the person's attributes such as name and address. Also like DNS, XDI was run by a single organization called \emph{XDI.org} that held a license to patents from the Cordance company.\footnote{\url{https://danbri.org/words/2008/01/29/266}} Although XRIs were put into early federated identity versions of OpenID~\cite{recordon2006openid}, XRIs failed to gain real-world usage, and they were eventually dropped from the more successful federated OpenID Connect system~\cite{openidconnect}, which was instead built instead on the IETF OAuth standard for authorizing accessing to data without globally unique identifiers~\cite{oauth2}. 

In contrast to the standards for digital identity that focus on assigning globally unique identifiers, cryptographic research focused on anonymous credentials that allowed users to directly show to verifiers their claims without revealing their identity, much less using a globally unique identifier~\cite{chaum1985security}. Of particular interest are zero-knowledge proofs for identity~\cite{camenisch2011framework} that led to attribute-based credential systems such as Microsoft's U-Prove~\cite{uprove} and IBM's Idemix~\cite{camenisch2002design} (currently used in Hyperledger Fabric.\footnote{\url{https://hyperledger-fabric.readthedocs.io/en/release-2.0/idemix.html}}) These anonymous credential schemes offered a high degree of privacy without trusted third parties. Although some schemes use blockchain technologies to achieve decentralization, these are still too computationally expensive for real-world use~\cite{garman2014decentralized}. There has even been some initial work like the SecureABC proposal on deploying cryptographic anonymous credentials in the context of immunity passports, although the ethical concerns still remain and there seems to be no move towards widespread implementation of the SecureABC anonymous credential scheme for immunity passports~\cite{hicks2020secureabc}.

One of the core problems with digital identity was the requirement for a centralized database of these globally unique identifiers. Blockchain technology appeared to both guarantee the non-collusion of identifiers and not require a centralized database of identifiers while enabling a seemingly infinite number of identifiers to be minted. Thus, the vision of a decentralized database of globally unique identifiers for people seems plausible technically, and  many of the efforts and people involved in prior work on digital identity re-emerged in the W3C in order to re-invent standards for a globally unique identity for every person on top of blockchain technology. This effort received relatively little attention until COVID-19 led to a push for immunity passports. While there earlier seemed no real use-case for a cross-border global identity system, immunity passports were seized upon as the `killer' use-case by groups like CCI. Given the rush to push for immunity passports and vaccination test results to revive travel by various governments, with trials even starting in the United Kingdom\footnote{The United Kingdom is testing the closed-source CommonPass by the Commons Project, as announced at \url{https://thecommonsproject.org/newsroom/safer-travel-and-accelerate-border-reopenings}} and funding for W3C digital identity standards by the Department of US Homeland Security,\footnote{\url{https://www.sbir.gov/sbirsearch/detail/1302459}} it would seem that a technical analysis of the W3C standards being pushed into these immunity passports is in order.

\section{W3C Verifiable Credentials}
\label{sec:vc}

Currently, all proposed immunity credential schemes rely on an obscure standard, the W3C Verified Credential Data Model 1.0 standard~\cite{w3c:vc}. A verified credential is defined by the specification as ``a tamper-evident credential that has authorship that can be cryptographically verified''~\cite{w3c:vc}, or in other words, a cryptographically signed message. However, rather than simply sign a byte-string, W3C Verifiable Credentials present an abstract data model for the idea of \emph{claims}, which are any list of attributes and values pertaining to a \emph{subject}, the ``entity about which claims are made''~\cite{w3c:did}. These claims are created by an \emph{issuer} who then creates a verifiable credential, which in turn are processed by a \emph{verifier}. The issuer is split into the issuer and the holder roles. For example, the issuer could be a medical laboratory that is testing the COVID-19 immune response of a patient, the subject and holder of a verifiable claim about their antibody status. While the specification is written to have one assume the holder is the person themselves, there is nothing to prevent the holder being a government database that the patient has no knowledge of.

More importantly, the claims are not a simple list of attribute value pairs or even arithmetic circuits that could be verified in zero-knowledge, but instead a graph build from the nearly forgotten W3C Semantic Web standards~\cite{berners2001semantic}, which has important ramifications upon processing. Namely, the Verified Credential specification recommends JSON or the Semantic Web serialization that uses JSON, JSON-LD~\cite{w3c:jsonld}. Cryptographic signatures can either be specified by IETF JSON Web Tokens~\cite{jwt} or the non-standardized ``Linked Data Proofs'' document~\cite{w3c:linkedproofs}. An example of a W3C Verifiable Credential for immunity credentials created by a member of the COVID Credentials Initiative is given in Figure \ref{fig:vc}.\footnote{\url{https://github.com/decentralized-identity/c19-vc.com/blob/master/src/bindingModels/ImmunoglobulinDetectionTestCard.json}}

\begin{figure}[th!]
\centering
\tiny
\begin{verbatim}
{
    "@context": [
        "https://www.w3.org/2018/credentials/v1",
        "https://w3c-ccg.github.io/vc-examples/covid-19/v1/v1.jsonld"
    ],
    "id": "http://example.com/credential/123",
    "type": [
        "VerifiableCredential",
        "ImmunoglobulinDetectionTestCard"
    ],
    "issuer": {
        "id": "did:web:vc.transmute.world",
        "location": {
            "@type": "CovidTestingFacility",
            "name": "Stanford Health Care",
            "url": "https://stanfordhealthcare.org/"
        }
    },
    "issuanceDate": "2019-12-11T03:50:55Z",
    "expirationDate": "2020-12-11T03:50:55Z",
    "name": "Immunoglobulin Detection Test Card",
    "description": "Immunoglobulin detection tests are based on the qualitative detection of IgM and IgG...",
    "credentialSubject": {
        "id": "did:key:z6MkjRagNiMu91DduvCvgEsqLZDVzrJzFrwahc4tXLt9DoHd",
        "type": "ImmunoglobulinDetectionTestSubject",
        "givenName": "Louis",
        "familyName": "Pasteur",
        "birthDate": "1958-07-17",
        "IgM": false,
        "IgG": true,
        "image": "data:...''
    }
}
\end{verbatim}
\caption{Example W3C Verified Credential for an ``Immunity Passport''}
\label{fig:vc}        
\end{figure}

\begin{figure}
\centering
 \includegraphics[width=.7\linewidth]{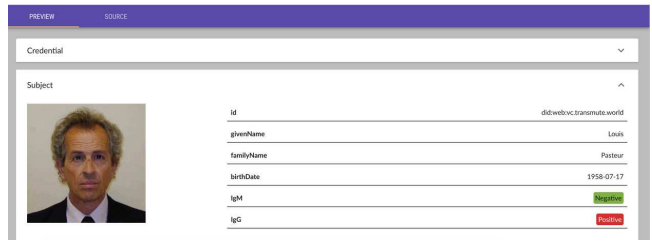}
\caption{Verifiable Presentation of an `Immunity Passport'}
\label{fig:graphic}
\end{figure}

As can be seen from this example, Verifiable Credentials essentially offer a number of mandatory properties, such as \texttt{type} to determine the kind of credential, an \texttt{id} (to refer to a W3C DID~\cite{w3c:did}), and \texttt{issuer} property with a date of issuance (\texttt{issuanceData}) and date of expiration (\texttt{expirationDate}), as well as information stored about the subject as  \texttt{credentialSubject}. Following W3C Semantic Web conventions used in the W3C Resource Description Framework (RDF) standard~\cite{w3c:rdf}, nearly all identifiers are identified either with a standard \texttt{https:} URL or a kind of DID~\cite{w3c:did}. As such a verifiable credential is simply a data format, the role of an application, as illustrated in Figure \ref{fig:graphic},\footnote{The image is from \url{https://github.com/decentralized-identity/c19-vc.com}} will be creating what is termed by the W3C as a \emph{verifiable presentation} that presents some of subset of the attributes for human inspection or further machine processing.

\textbf{Critique.}  Note that injectivity of the serialization scheme is necessary for the security of digital signatures. The W3C Verifiable Credentials standard can depend on the problematic Semantic Web RDF format, which lacks standardized bit-serialization necessary for signatures~\cite{signingrdf}. RDF does not specify a syntax like XML, but `semantic' graphs of URIs  (Uniform Resource Identifiers, such as \emph{http://example.org}) and values where the same graph can be serialized in different manners~\cite{w3c:rdf}. Worse, there is no unified way to skolemize `blank nodes' (i.e. existential variables) across serialization formats. Instead, RDF features a number of ad-hoc canonicalization forms given in non-standard documents, with W3C VC relying on a non-standard algorithm.\footnote{The ``RDF Dataset Normalization'' document \url{http://json-ld.github.io/normalization/spec/}}  Also, Semantic Web serialization can depend on the resolution of external documents to URIs to `link' data, similar in spirit to XML namespaces~\cite{w3c:rdf}. The Semantic Web also even has issues with TLS support~\cite{halpin:semanticinsecurity}. Further increasing the likelihood of attacks, implementers are recommended that Linked Data Proofs ``are detached from the actual payload.''\footnote{\url{https://www.w3.org/TR/vc-imp-guide/}} The combination of these problems, where a variable number of  signatures can be arbitrarily detached and re-attached to messages combined with the (possibly insecure) retrieving of unknown external documents leads RDF in general, and likely Verifiable Credential implementations that use RDF, to be vulnerable to the same kind of attacks that rendered XML digital signatures insecure in practice~\cite{mcintosh2005xml}.

In detail, this leads to \emph{signature exclusion} and \emph{signature replacement} attacks where an adversary can remove the signature of a signed message, perhaps replace it with another signature,  and to trick the verifier into falsely accepting the  message as valid due to ambiguity in parsing. There exists a mitigation of these canonicalization and serialization issues as VCs can be used without RDF, and instead serialized with the well-specified IETF JSON Web Tokens (JWT) serialization~\cite{jwt}. Yet even in this case, there still exist in some JWT implementations (although the IETF specifications have fixed these issues in the specification) the ability to easily misuse cryptography, which can lead to intentional \emph{cryptographic downgrade} attacks on the cryptography in JWTs~\cite{jager2013one}. In general, it seems as if all that is needed is to prove a signature of a valid health care provider or hospital on a credential presented by a user; it is unclear why an error-prone data format without a clear byte order that is badly suited for cryptography should be used. A simple signature over a byte-string signifying the result would suffice.

The most likely reason for using Semantic Web standards is that W3C Verifiable Credentials are to be used for data integration rather than privacy. Semantic Web standards are notoriously inefficient due to being stored as labeled graphs with existential variables as well as cycles, making canonicalization difficult~\cite{groppe2011data}. Each blank node must be uniformly given a stable identifier, and different graphs can be output in different orders. Even if canonicalized and then serialized, the graph isomorphism problem is NP complete so it cannot be determined if two graphs match efficiently. The Semantic Web has found usage in government data integration by the Department of Homeland Security~\cite{ding2005homeland} as the goal of the W3C standards is to allow the ``linking'' of data via the reuse of URIs as labels of nodes in the graph~\cite{w3c:jsonld}. Linking of user data seems at odds with user  privacy, as privacy is typically defined as unlinkability~\cite{pfitzmann2010terminology}. However, the ability to link patient data may have use cases for the government and medical data.

\section{W3C Decentralized Identifiers}
\label{sec:did}

Decentralized identity management means that identity is provisioned using a decentralized technical infrastructure such as distributed ledger technology~\cite{dunphy2018first}, and so there is no single identity provider or key registry. A \textbf{Decentralized Identifier (DID)} is a W3C standard under development to provide a uniform naming scheme for what is termed ``subjects'' to be identified~\cite{w3c:did}. The scope of subjects has cosmic ambitions: ``Anything can be a DID subject: person, group, organization, physical thing, digital thing, logical thing'' but the DID document maintains that each DID is unique per subject, so that ``a DID has exactly one DID subject'' and that somehow ``a DID is bound exclusively and permanently to its one and only subject''~\cite{w3c:did}. In the context of immunity credentials, a DID is an attempt to create one or more universally unique identifiers for every person that includes whether or not they have been infected by COVID-19 or not.

Technically, a DID is simply a URI scheme that specifies a \emph{DID method}, which in turn is used to resolve a DID into a concrete (if possibly unsigned) \emph{DID document}.  The \emph{DID document} in turn contains references to key material and routing information to one or more \emph{DID service endpoints} that allow access to yet another document describing the subject of the DID and so including explicitly personally identifiable information (PII) in RDF using VCs. DIDs are represented as an \texttt{id} relationship in a DID document. An example DID is shown in Figure \ref{fig:vc} as \texttt{did:key:z6MkjRagNiMu91DduvCvgEsqLZDVzrJzFrwahc4tXLt9DoHd}. This method, as defined in yet another non-standardized document,\footnote{\url{https://w3c-ccg.github.io/did-method-key/}} defines a DID document where a single Ed25519 public key is used for signing documents as well as authentication. More complex DID methods exist and each new DID-enabled system can add register a new method.\footnote{\url{https://w3c.github.io/did-spec-registries/}}

For example, another kind of DID could resolve to a DID document via access to a blockchain such as the DID-centric Sovrin blockchain.\footnote{\url{https://sovrin-foundation.github.io/sovrin/spec/did-method-spec-template.html}} For example, the DID \texttt{did:sov:29wksjcn38djfh47ruqrtcd5} retrieves a DID document from the Sovrin blockchain,\footnote{\url{https://sovrin.org/library/sovrin-protocol-and-token-white-paper/}} which is considered a \emph{verifiable registry} as much as a centralized server or the Bitcoin blockchain according to the W3C Verifiable Credentials standard~\cite{w3c:vc}. Verifiable registries are simply a public key infrastructure for the key discovery needed to verify a W3C VC, and so a blockchain is used in this context merely as a global public database of time-stamped DID documents that contain keys and service end-points. An end-point can retrieve additional personal data about DID subjects like an immunity credential.

\textbf{Critique.}  Although the idea of a registry with at least one unique identifier for every possible object that may exist could be considered itself a suspect concept for a standard, W3C DIDs suffer from a number of purely technical weaknesses. The resolution from a DID to a DID document is customized per implementation, and in practice permissioned federations such as Sovrin resolve the DID to the necessary DID document, and so the blockchain is equivalent to using a public database of DID documents (without access control) replicated by a threshold number of trusted authorities. In other words, there seems to be no technical reason outside availability to use a blockchain rather than a trusted public third party database for the verifiable registry containing DIDs~\cite{wust2018you}. Unlike other blockchain-based systems like Claimchain~\cite{claimchain}, there is no access control specified and all identifiers are shared in public.  The privacy issues are to some extent worse in W3C DIDs than in even centralized or federated identity systems, as correlation attacks may not only be done by a malicious identity provider, but by any actor as typical DIDs are stored in public chains and, if the DID document is public, anyone can search the chain to discover the times of changes not only to DIDs, but to key material if the DID document is public.\footnote{Although the DID standard permits these attacks, these issues could in theory be addressed per implementation in a customized manner, as Microsoft claims is done in their ION identity system: \url{https://techcommunity.microsoft.com/t5/identity-standards-blog/ion-booting-up-the-network/ba-p/1441552}}

Most importantly, there is no way to enforce the privacy of any attributes attached to the Verifiable Credentials produced by service end-points accessed via DID documents and DIDs. As for DID documents themselves, it is only stated that ``it is strongly recommended that DID documents contain no PII''~\cite{w3c:did}. As they are effectively permanent identifiers written to a public blockchain, W3C DIDs may of course be correlated. One-time use DIDs are not enough, as the W3C DID standard notes ``the anti-correlation protections of pseudonymous DIDs are easily defeated if the data in the corresponding DID Documents can be correlated''~\cite{w3c:did}. While it could be claimed that simply storing key material and end-points for services does not compromise user privacy and only identifiers without personal data is stored on the chain, storing identifiers as well as keys and service end-points publically can leak valuable data. For example, the addition of service end-points for COVID-19 testing centers would leak the fact that a person with this key material had been tested for COVID-19, along with the likely approximate physical geolocation as correlated with the hospital and day of the test, as given by the update to the DID document.

Although DIDs claim to support zero-knowledge proofs, there is no advanced cryptography used outside of RSA and elliptic curve Ed25519 signatures in the standards themselves~\cite{w3c:did}. In other publications, at least one of the authors of the W3C DID standard believes a ``trusted witness,'' i.e. a trusted third party nominated by the user with control over their personal data, is a better paradigm than zero-knowledge proofs due to the lack of trust in the verifier or holder of a credential~\cite{arnold2019zero}, although the solution of adding yet another trusted third party does not seem to solve the actual problem, but merely displace it. Other DID deployments, like those at Microsoft, claim to that they will use zero-knowledge proofs but end up being anonymous credentials with no special need for the identity machinery of DIDs and VCs.\footnote{\url{https://github.com/decentralized-identity/snark-credentials/blob/master/whitepaper.pdf}}

Another question is whether or not the blockchain is necessary at all.  If DIDs are indeed supposed to be for one-time usage, then why use DIDs on a blockchain at all rather than just an anonymous credential scheme without DIDs? The actual Verifiable Credentials are not stored on the blockchain but stored by service end-points, which can be centralized servers rather than some necessarily decentralized actor. Indeed, it appears that in many of these cases, it is assumed the actual party that holds the VC document is a third-party service that functions as an \emph{identity provider}. This identity provider could be relatively benign, such as the hospital that performed the test, but it could be yet another for-profit blockchain-based digital identity startup. Rather than disintermediate identity providers like Facebook or government identity databases, the actual database access simply requires the additional step of contacting yet another third party database, namely the verifiable registry, before communicating with their actual personal data pointed to from the verifiable registry. The blockchains used as verifiable registries can be decentralized but can also be federated permissioned blockchains using ``proof of authority'' where a quorum, or sometimes any member, can simply add a block with a new DID identifier. So not only is there no gain in terms of privacy, but no gain in terms of decentralization if decentralization is to be defined as the absence of a trusted component in a system~\cite{troncoso2017systematizing}. A system would leak far less data if the holder of a VC simply included a method to directly contact the issuer with the VC itself rather than use excessive redirection based on a public blockchain for key management.

\section{Immunity Credential Systems: A Case Study}
\label{sec:app}

Although there are many systems for immunity credentials being built (currently over fifteen) in countries ranging from Hong Kong to Italy according to the COVID-19 Credential Initiative Implementation Workstream Homepage,\footnote{\url{https://docs.google.com/document/d/1dbWvs1m8uziTsbhUQv_nPofTXAyDSkxI5CZtoo1SlRY}} most of the published proposals are high-level sketches~\cite{bansal2020optimizing}.  As of late 2020, only one academically published immunity credential system has a detailed specification called ``COVID-19 Antibody Test/Vaccination Certification: There's an App For That,'' by Eisenstadt et al.~\cite{eisenstadt}. Like the CCI work, this app uses the DID and VC W3C standards. The paper puts forward the technical design goals that an immunity credential is meant to address in addition to the non-technical requirements of being ``cost-effective'' and ``easy to administer''~\cite{eisenstadt}:

  \begin{itemize}

  \item \textbf{Privacy-preserving} 
  \item \textbf{Un-forgeable}
  \item \textbf{Scalable}: ``to millions of users'' 
  \item \textbf{Easily verifiable}: ``while still preserving privacy''

  \end{itemize}

  The overall motivation for using digital certificates for immunity passports is that, not only are they more cost-effective and more scalable but that they can be secure as ``a paper version is too vulnerable to alteration or forgery''~\cite{eisenstadt}. It seems as if the primary security argument of Eisenstadt et al. is that digital signatures are unforgable, and so could prevent both a hospital from forging a credential or a person from using a false credential. They also seemingly narrowly conceive of privacy as only the explicit prevention of the release of personal identifying information such as a patient's name and social security number, while traditionally it would be considered that linking usages of a credential would violate privacy. There is no explicit threat model given, but it does seem like the authors are aiming for a weak local passive adversary, rather than an active adversary that can observe changes of state in the blockchain used by DIDs in order to re-identify subjects or one that can carry out replay attacks with credentials.

  The minimal PII needed by each actor in the immunity credential system is not detailed.  The privacy and security properties are claimed to be fulfilled by the use of W3C DIDs in tandem with the actual VCs containing the immunity results stored using a Semantic Web architecture known as `Solid,' designed by W3C director and inventor of the Web, Berners-Lee~\cite{mansour2016demonstration}. This architecture is composed of \emph{pods} which store RDF data like VCs using HTTP access points. The core concept is that personal data can be stored locally on a device like a mobile phone or even a ``favorite cloud server''~\cite{eisenstadt}.  The claim of Eisendstat et al. is that ``the provider's access to the data is limited by the user's preference,'' although currently it appears no cryptographic techniques are used to encrypt the data in Solid~\cite{eisenstadt}. There is an unformalized access control language used by pods.\footnote{\url{https://github.com/solid/web-access-control-spec}} Currently backed by a startup called Inrupt,\footnote{\url{https://inrupt.com/}} the platform is build in Javascript.


  The app uses a distributed ledger called OpenEthereum, a fork of Ethereum by the Open University and ran by a consortium (and so is call a ``Consortium blockchain''). In contrast to Ethereum but similar to other DID-based chains like Sovrin, it is based on ``proof of authority'' (i.e. a permissioned blockchain where any validator or quorum of validators may write to the chain, but not other actors like users). It is then argued that while DIDs allow pseudonyms and Solid pods give an user ``a choice regarding whether and where to host personal information'' (which would include the results of the COVID-19 tests), ``a hash of the Verifiable Credential is stored on the Consortium blockchain''so that the immunity credential itself can be verified~\cite{eisenstadt}.

\begin{figure}
\centering
 \includegraphics[width=.8\linewidth]{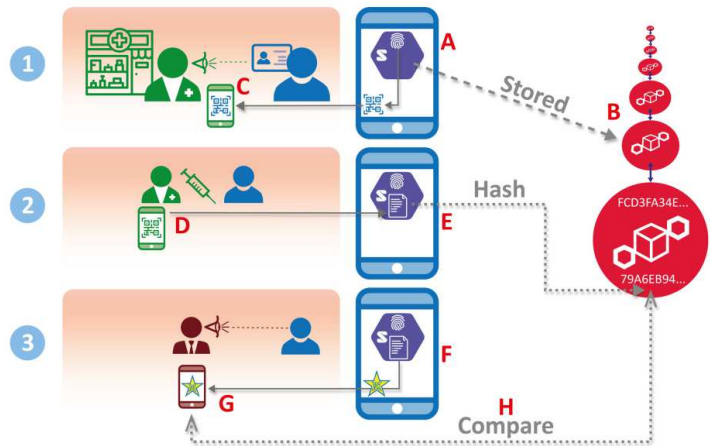}
\caption{Information Flow of an W3C Verified Credential and DID for an ``immunity passport'' as given in Eisenstadt et al.~\cite{eisenstadt} }
\label{fig:app}
\end{figure}

  The information flow of the application is given below and illustrated in Figure \ref{fig:app}, where information is assumed to shared using QR codes:
  \begin{itemize}
  \item \textbf{Step 1.}  The patient (the holder, also the ``subject'' of a DID) and the hospital doing the immunoglobulin tests (the issuer) are assigned DIDs (A) that can be resolved on the permissioned blockchain (B).  After an identity check (C), the patient gives their DID and required personal information for an immunity credential to the hospital.
  \item \textbf{Step 2.} The hospital gives antibody test (D)  and sends its DID with the test results signed by its public key to the patient (E), which then receives it and uploads a hash of the signed immunity credential to the blockchain. 
  \item \textbf{Steps 3.} The patient can then show their employer (the verifier) the immunity credential (F) with associated DID (G), who can then check the hash of the credential by looking up the DID and verifying if the signature of the test results in the immunity credential are correctly attributed to the hospital by retrieving the hospital's DID and retrieving the hash of immunity credential from the blockchain to compare it to the immunity credential shown by the patient (H), and then approve or deny the patient some action. 
  \end{itemize}

\textbf{Critique}: As Solid is being backed by the inventor of the Web, one would hope it would offer privacy or security properties. Yet Solid offers no security properties or privacy properties, currently having only the aforementioned access control language but usage of kernel security modules or other ways of providing assurance are missing. There are draft ideas on authentication.\footnote{\url{https://solid.github.io/authentication-panel/solid-oidc/}} It seems that giving a user the choice of where to store their COVID-19 antibody test results will likely not lead to more security, as  users may store their test results on a mobile phone whose operating system needs updating, a insecure personal server, or in some other location that is highly unlikely to be secure in terms of systems administration. While Solid democratizes storage and avoids creating a centralized honey-pot of test results, does it make sense to allow people to be given a technical choice over where to store their data rather than embedded in the operating system or using secure enclaves? The use of RDF, due to the  underlying graph model, has poor scalability compared to traditional relational databases or key-value stores~\cite{groppe2011data}. The main purpose of VCs using RDF in Solid appears to be to integrate data using unique identifiers based on URIs like DIDs.

The use of Semantic Web technologies appears to be designed to make medical data more amendable to fusion and analytics by hospitals or governments, given there is a dearth of user-centric services that consume or produce Semantic Web data. This makes it difficult to believe that ``pairwise-unique DIDs and public keys'' will be used in practice, as that would eliminate any benefits of using Semantic Web techniques~\cite{eisenstadt}. If a DID was used just once, then why have an identifier at all? Another core idea is that ``user preference'' can also eliminate the capture of data by servers, as Eisenstadt et al. claim that ``Everything in this app is decentralized. Anyone wishing to abandon involvement in this kind of certification can just delete the Verifiable Credentials stored on their Solid Pods. There will be no records whatsoever, as if they had never been on the system''~\cite{eisenstadt}.  Yet there is no serious proposal to prevent the operating system of a mobile phone or a server from copying the immunity certificate for the purposes of sharing, and deletion cannot be guaranteed to have actually happened.

  The authors of the paper also seem to have misconceptions about cryptography, leading to insecure uses of hashes for storing immunity credential information in a publicly accessible blockchain. In particular, the authors hold that storing the hash of their immunity certificate on the blockchain allows ``individuals who have been tested to change their minds and quit the scheme, knowing that even cryptographically encoded data will be ‘orphaned’ (no data pointing to it), rendering it meaningless''~\cite{eisenstadt}. The authors seem not to be aware of dictionary attacks and how the storage of sensitive data using hashes can be securely implemented. Although it is true that for arbitrary data, a hash is a one-way function that can not be reversed per se and maps this data to fixed size values, by itself a hash function is typically deterministic and the possible claims embedded in the credential are finite. For example, the number of unique birth-dates of living individuals is on the order of fifty thousand, and so if one wishes one to discover a hashed birthdate, one simply iterates through the birthday values (as in a `dictionary')  with the hash function until a match is discovered. This would be even simpler for test results that are either positive or negative, and continues to holds true even when ``deleting data on the Solid Pods'' although the authors claim that this ``will also turn the hashes on the blockchain into ‘orphans’ (no data pointing to the hash), i.e. the hashes will become meaningless: it is not possible to recover the original data from a hash''~\cite{eisenstadt}. If an adversary knows some fairly basic personal data about a person such as their name and birthday, as per the example in Figure \ref{fig:vc}, and one wished to discover their test results, an adversary could simply iterate through the possible results of an immunity test to determine the results of a patient's immunity test from only the hash on the blockchain.

  The storage of even hashed immunity certificates on a public blockchain also is a poorly thought out idea for avoiding leaking sensitive data on a public blockchain. Using a hash would only make sense if appropriate salting could be done, but the salt would have to be somehow shared to verify the hash. Simply using a seemingly `random' string of bytes in the VC like a photo would not be enough to secure the hash, as the photo would be revealed and remain the same over usages of the credential and so would not be a salt, as an adversary could capture it.  Alternatively, publishing the ciphertext of the results encrypted under an ephemeral asymmetric public key on the blockchain, where the corresponding decryption key could be given on an as needed basis, would be a better design. Nonetheless, having such sensitive data permanently available on a public blockchain, even if encrypted, would be a risk. W3C DIDs attempt to do the right thing by publishing only a reference to a key on chain. Even in the case where only a public key is published on the blockchain, it would be better to use encryption than hashing to preserve privacy, as has been explored in work with well-defined privacy and security properties for decentralized key management~\cite{claimchain}.

  Given these earlier problems, it should not come as a surprise that this immunity credential architecture does not address the fundamental problem of digital immunity passports: Having this information in digital form by nature increases the ability for this data to be copied and altered. Although the use of cryptography can prevent some of these attacks, this is not the case in the proposed architecture. For example, the verifier could not copy the passport and do a replay attack with the credential. This is easily possible, as no private key material operations are required for usage of the immunity credential that require the holder having any secret key material such as decryption of or selectively disclosing the credential.

  Surprisingly, the only cryptographic operation the verifier has to do is retrieve the public key of the issuer from the blockchain. Despite various claims that ``the app allows the user selectively to present only the specific test result,'' no details on zero-knowledge proofs are given by Eisenstadt et al.~\cite{eisenstadt}.  It is recommended by Eisenstadt et al. that when using digital credentials an identity check with a physical photographic identity card is done, although this methodology would also argue in favor of simply using physical paper certificates as immunity credentials~\cite{eisenstadt}. Eisenstadt et al. propose that it should be possible to have ``burned'' the photographic identity into the credential, i.e. have a copy of a valid visual identity as part of the credential so that physical national identity paperwork should not have to be checked when showing an immunity credential~\cite{eisenstadt}, although this increases the amount of sensitive data used in the credential. In terms of W3C standardization, methods for verifying a credential holder's identity is left outside of the scope of the document. There is a mention of a challenge-response protocol to prevent replay attacks given in the non-normative Verifiable Credentials Implementation Guidelines but the protocol, as outlined, uses an unsigned nonce and so is vulnerable to replay attacks.\footnote{\url{https://www.w3.org/TR/vc-imp-guide/}}

  The privacy properties of the immunity passport scheme proposed are claimed to result primarily from the supposed virtues of the underlying W3C standards ``the concepts underlying Verifiable Credentials and the Decentralized Verification of Data with Confidentiality are diametrically opposed to any kind of central data storage or Big Brother-style snooping and data collection, and indeed provide excellent and agreed standards for avoiding such snooping and data collection''~\cite{eisenstadt}. As shown earlier, these claims over the privay and security of the W3C DID and VC standards are dubious. The immunity passport design put forward by Eisenstadt et al. seems aimed at those with a background in the ``Semantic Web'' (a cluster of W3C standards for data management~\cite{hepp2005semantic}) and are missing cryptographic security assumptions as well as a realistic privacy impact assessment. However, there may be other immunity credential schemes that address these concerns in the future that may be attempted to be rolled out in the future. The problem is both the idea of immunity credentials and the standards used to implement, and these are separable if interlinked problems. Trying to move the paradigm from the ill-conceived idea of immunity passports to the newer but very similar idea of vaccination test certification, Eisenstadt et al. (whose earlier versions\footnote{\url{https://arxiv.org/abs/2004.07376}} envisioned deployment to the whole population) now state that credentials ``should only be applied to workers in healthcare and other comparable key sectors'' and added the term ``vaccination'' to their title~\cite{eisenstadt}. Yet the design still has the same problems. The other problem is that there is a more structural issue regarding how these W3C standards like VCs and DIDs came to become standardized and therefore assumed to be suitable for high assurance use-cases like immunity credentials.

\section{The Abuse of Security Standardization}
\label{sec:abuse}

The goal of security standardization should be both to guarantee the security and privacy properties of a particular technology and then promote their widest deployment. The technical proposals for immunity passports have almost all entirely been based on the World Wide Web Consortium's Decentralized Identifier (W3C DID) and Verifiable Credentials (W3C VC) standards, and these standards are currently being proposed for widespread usage by groups like the COVID-19 Credentials Initiative. The problem is two-fold: 1) a lack of clarity on what is a standard at the W3C and 2) the lack of review by the wider security and privacy community despite being standardized, in the case of W3C VC, or ``standards-track'' like W3C DID. A diagnosis of the underlying issues is required in order to assure the high quality of future security standardization.

First, what makes a standard? In the case of the W3C, a standard is a standard by virtue of a guarantee of royalty-free licensing of the underlying technology. In the case of the growth of the Web, the importance of the W3C is that it is in effect a patent pool for the World Wide Web that allowed many developers and companies to build on the Web in a permission-less fashion.\footnote{Note that a patent holder can still claim patent infringement even if an idea is embodied in a standard (such as an IETF RFC) and in open source code.} W3C standards are explicitly licensed by W3C members under a royalty-free license.\footnote{\url{https://www.w3.org/Consortium/Patent-Policy-20040205/}} In contrast, the IETF ``Note Well'' policy simply requires disclosure of known patents by individuals.\footnote{\url{http://www.rfc-editor.org/rfc/rfc3979.txt}} The much stronger W3C policy creates a kind of `patent war-chest' composed of all W3C standards, from XML to HTML5. This patent pool is then enforced by a `balance of terror' so that any member that makes a patent claim on a W3C standard triggers their loss of royalty-free licensing for \emph{all} W3C standards. The W3C patent pool was created precisely as an attempt to prevent patents from becoming part of standards at the W3C. The membership requirement of the W3C, as opposed to the more informal IETF, is due in part to the licensing requirements of patents. This protection against patents is one of the primary features of standards as a common good.

The problem is that the line between what is a standard and what is not a standard at the W3C has blurred.  Over the years the relative importance of patents declined and the importance of communities based on open standards and protocols increased, first due to the development of HTML as a `living standard' by WHATWG outside of the W3C,\footnote{\url{https://html.spec.whatwg.org/}} and then as shown by the explosion of interest in blockchain technologies since 2017. This led to the formation at the W3C of Community Groups, which are open to all and have no review from the rest of the W3C but can produce documents that appear to have the imprimcater of the W3C without review, with patent licensing being only to contributions and only opt-in to the entire specification.

In the case of W3C's digital identity efforts around W3C VCs and DIDs, there is a larger extended group of documents, ranging from the nascent DID Authentication to Linked Data Proofs, that it appears are also crucial to immunity passports and DID usage in general. These documents originate in either a W3C Community Group  or the newly created Decentralized Identity Foundation,\footnote{\url{https://identity.foundation}} but are considered as ``standards'' by proponents of W3C Verifiable Credentials and DIDs. This could confuse anyone, including a government interested in immunity passports, into thinking the underlying technology was both unburdened of intellectual property and of a high standard in terms of security and privacy. Furthermore, this development of an endless multitude of non-standards (mostly by the same small editorial group as the W3C VC and DID standards) makes implementation by developers an error-prone work of endless exegesis and seems to serve primarily to deflect criticism from the actual problems with the existing W3C standards for digital identity. Every privacy and security issue with DIDs and VCs can be claimed to be `solved' by yet another half-baked non-standard document or a product that has some yet-to-be-released non-standardized feature. Therefore, standards bodies should clearly separate standards from non-standards, scope standardization efforts to a finite number of documents, and reduce the dependency of the former on the latter.

Second, the purpose of security standardization can be construed as wide review and analysis of the security properties of standards. It is unclear how Verifiable Credentials and DIDs became W3C standards without review by security and privacy experts. One reason could be the influx of dues-paying blockchain companies as W3C members and a lack of attention from browser-oriented companies. As Verifiable Credentials were relatively unsuccessful in a large number of fields, ranging from internet of things~\cite{bartolomeu2019self} to educational credentials~\cite{kontzinos2020decentralised}, the work did not attract the attention of the security research community. Also, as the standardization effort has no clear boundaries, it is unclear where to begin or end such a security and privacy review. With the growing interest from governments in immunity passport schemes, the security community should focus on these global identity schemes. Yet these identity standards have crucial dependencies on non-standardized documents that are in a state of flux.

The W3C and other standards bodies should impose more stringent guidelines for security and privacy review on their future work. Overall, although the group of documents needed for W3C VC-based immunity passports is vast, these documents do not possess the technical (and in particular, cryptographic) detail needed for a security and privacy analysis, much less formal verification. Our thesis is that the underlying effort around immunity passports and associated W3C efforts are examples of \emph{cryptography theater}, which we define as the appeal to cryptography without a concrete specification or protocol in order to claim to be secure without necessarily being so.  While the designers of W3C DIDs and VCs may have designed the technology to the best of their ability, every standard should have a rigorous security review. A state of emergency caused by COVID-19 should require more, not less, review in terms of security and privacy for immunity passports given the possibility of widespread usage and abuse.



\section{Conclusion}
\label{sec:con}

Our vision of security standardization is that it should either provide wide review of proposed standards in order to correct security and privacy flaws, or prevent their standardization in the first place. The fact that specifications like W3C Verifiable Credentials even became standards is problematic without security and privacy review by experts, and it would be better if future standards that touched upon security and privacy were done at the IETF and follow a more rigorous multi-stakeholder process that involves academics and verification of the claims of security properties of the standards, as was done in TLS 1.3~\cite{bhargavan2017verified}. The unnecessary complexity and lack of review of these standards can lead to concrete privacy and security harms for users, who are naturally confused by claims of privacy and security dependent on cryptography.

The use of W3C standards to legitimize immunity passports is a prime example of how a security standardization process, without an actual functional standards body that achieves wide review by experts, can be hijacked by self-interested government or business interests without providing any protection for users.  It simply is dangerous to build on standards that are not well-understood, and standards bodies like the W3C that `rubber stamp' such standards should be held to account by the security research community.

As shown, there may be concrete steps that can help the W3C. For example,  all RDF related formats can be dropped from VCs until bit-serialization is standardized. Work on DIDs at the W3C can simply be halted as global identity on a blockchain cannot be done in a secure and privacy-preserving manner without advanced cryptography, which DIDs lack. For the use-cases of identity management, there are many well-known cryptographic techniques that offer strong and rigorous guarantees of privacy and security, although they are not used as the foundation of the W3C standards for digital identity, but merely included as an optional afterthought. One somewhat surprising aspect of immunity passport proposals is their reliance on blockchain technology. Blockchain technology has uses, but these uses should be justified in terms of concrete security and availability properties given by design goals.

The most concrete immunity passport proposal dangerously puts the hash of personal data on the  blockchain~\cite{eisenstadt}.  Even the use of blockchain technology by specifying resolution of an on-chain mapping of an identifier to a key in systems like Sovrin ends up being a redirect to centralized servers, undermining a claim of the blockchain promoting decentralization. As the use of blockchain technology does not seem necessary for the goals of the immunity passports and likely hinders rather than helps privacy, immunity passports -- and more widely both W3C DIDs and VCs -- use blockchain for blockchain's sake.\footnote{One plausible reason for a blockchain would be censorship-resistance via peer-to-peer gossip networks, but this seems to be an implausible goal for immunity credentials. Lastly, a decentralized PKI in of itself does not require a blockchain even for censorship resistance of key material~\cite{claimchain}.} 

The problem is not just one of broken standards and an off-the-rails standardization process at the W3C. The conflict over DRM at the W3C demonstrated already the W3C standards process was prone for corporate capture and capable of being abused~\cite{halpin:eme}. The underlying problem is the cultish desire for a ``self-sovereign'' global identity system runs counter to privacy. Standards bodies should avoid cults. The technical proposal for immunity passports based on W3C DIDs could allow the COVID-19 crisis to be a driver for a larger vision of a global digital identity system where every single human has a permanent and globally unique identifier. This form of digital identity runs counter to privacy, opening the door for a new form of \emph{identity totalitarianism} where every person must be identified in a database -- of which a blockchain is merely a fashionable new form -- in order to be part of society.

The question should not be whether or not immunity passports can be technically secure and private, but whether or not they should be built at all. Due to the state of emergency caused by COVID-19, fundamental rights -- such as the freedom of movement -- could be taken away based on data connected to persistent digital identity. Yet temporary measures meant for a purpose as seemingly harmless as reviving tourism could become normalized as the blockchain-based identity databases are by design permanent and are difficult to disassemble once the crisis has past. Blockchain technology could just as easily allow automated discrimination based on personal data as it could enable travel during COVID-19, and form the technical basis for a `social credit' system that crosses borders. 

Digital identity has many use-cases outside of immunity passports, from the relatively benign domain of education to the critical infrastructure of medical data, but also many dangers. For example, the use of blockchain identities for refugees could be useful in allowing them access to bank accounts, but could also be easily used for surveillance.\footnote{\url{https://www.qeh.ox.ac.uk/content/blockchain-refugees-great-hopes-deep-concerns}} Identity systems exist to help large institutions manage and control populations.  The promotion of digital immunity passports using the rhetoric of decentralization and self-sovereignty may be appealing and done with the best of intentions, but the COVID-19 crisis should not be treated as an excuse to push out standards or software that may harm users. 

\bibliographystyle{plain}
\bibliography{main}

\begin{thebibliography}{10}

\bibitem{arnold2019zero}
Rachel Arnold and Dave Longley.
\newblock Zero-knowledge proofs do not solve the privacy-trust problem of
  attribute-based credentials: {W}hat if {A}lice is evil?
\newblock {\em IEEE Communications Standards Magazine}, 3(4):26--31, 2019.

\bibitem{bansal2020optimizing}
Agam Bansal, Chandan Garg, and Rana~P Padappayil.
\newblock Optimizing the implementation of {COVID}-19 {I}mmunity {C}ertificates
  using blockchain.
\newblock {\em Journal of Medical Systems}, 44(9):1--2, 2020.

\bibitem{bartolomeu2019self}
Paulo~C Bartolomeu, Emanuel Vieira, Seyed~M Hosseini, and Joaquim Ferreira.
\newblock Self-sovereign identity: Use-cases, technologies, and challenges for
  industrial iot.
\newblock In {\em 2019 24th IEEE International Conference on Emerging
  Technologies and Factory Automation (ETFA)}, pages 1173--1180. IEEE, 2019.

\bibitem{berners2001semantic}
Tim Berners-Lee, James Hendler, and Ora Lassila.
\newblock The {S}emantic {W}eb.
\newblock {\em Scientific american}, 284(5):34--43, 2001.

\bibitem{bhargavan2017verified}
Karthikeyan Bhargavan, Bruno Blanchet, and Nadim Kobeissi.
\newblock Verified models and reference implementations for the {TLS} 1.3
  standard candidate.
\newblock In {\em Security and Privacy (SP), 2017 IEEE Symposium on}, pages
  483--502. IEEE, 2017.

\bibitem{uprove}
S~Brands and C~Paquin.
\newblock U-{P}rove cryptographic specification v1.0, 2010.

\bibitem{camenisch2011framework}
Jan Camenisch, Stephan Krenn, and Victor Shoup.
\newblock A framework for practical universally composable zero-knowledge
  protocols.
\newblock In {\em International Conference on the Theory and Application of
  Cryptology and Information Security}, pages 449--467. Springer, 2011.

\bibitem{camenisch2002design}
Jan Camenisch and Els Van~Herreweghen.
\newblock Design and implementation of the {I}demix anonymous credential
  system.
\newblock In {\em Proceedings of the 9th ACM conference on Computer and
  communications security}, pages 21--30, 2002.

\bibitem{signingrdf}
Jeremy~J Carroll.
\newblock Signing {RDF} graphs.
\newblock In {\em International Semantic Web Conference}, pages 369--384.
  Springer, 2003.

\bibitem{chaum1985security}
David Chaum.
\newblock Security without identification: Transaction systems to make {B}ig
  {B}rother obsolete.
\newblock {\em Communications of the ACM}, 28(10):1030--1044, 1985.

\bibitem{ding2005homeland}
Li~Ding, Pranam Kolari, Tim Finin, Anupam Joshi, Yun Peng, Yelena Yesha, et~al.
\newblock On homeland security and the {S}emantic {W}eb: A provenance and trust
  aware inference framework.
\newblock In {\em Proceedings of the AAAI Spring Symposium on AI Technologies
  for Homeland Security}, 2005.

\bibitem{dunphy2018first}
Paul Dunphy and Fabien~AP Petitcolas.
\newblock A first look at identity management schemes on the blockchain.
\newblock {\em IEEE Security \& Privacy}, 16(4):20--29, 2018.

\bibitem{eisenstadt}
Marc Eisenstadt, Manoharan Ramachandran, Niaz Chowdhury, Allan Third, and John
  Domingue.
\newblock {COVID}-19 antibody test certification: {T}here's an app for that.
\newblock {\em IEEE Open Journal of Engineering in Medicine and Biology},
  1:148--155, 2020.

\bibitem{garman2014decentralized}
Christina Garman, Matthew Green, and Ian Miers.
\newblock Decentralized anonymous credentials.
\newblock In {\em Proceedings of the {N}etwork and {D}istributed {S}ystem
  {S}ecurity {S}ymposium - {NDSS}'14}. Internet Society, February 2014.

\bibitem{groppe2011data}
Sven Groppe.
\newblock {\em Data management and query processing in semantic web databases}.
\newblock Springer Science \& Business Media, 2011.

\bibitem{halpin:eme}
Harry Halpin.
\newblock The crisis of standardizing {DRM}: The case of {W3C} {E}ncrypted
  {M}edia {E}xtensions.
\newblock In {\em International Conference on Security, Privacy, and Applied
  Cryptography Engineering}, pages 10--29. Springer, 2017.

\bibitem{halpin:semanticinsecurity}
Harry Halpin.
\newblock Semantic {I}nsecurity: {S}ecurity and the {S}emantic {W}eb.
\newblock In {\em Society, Privacy and the Semantic Web-Policy and Technology
  (PrivOn 2017)}, 2017.

\bibitem{halpin:standards}
Harry Halpin.
\newblock Decentralizing the social web.
\newblock In {\em International Conference on Internet Science}, pages
  187--202. Springer, 2018.

\bibitem{oauth2}
Dick Hardt.
\newblock The {OA}uth 2.0 authorization framework.
\newblock {\em IETF RFC 6749}, 2012.
\newblock \url{https://tools.ietf.org/html/rfc6749}.

\bibitem{hepp2005semantic}
Martin Hepp, Frank Leymann, John Domingue, Alexander Wahler, and Dieter Fensel.
\newblock Semantic business process management: A vision towards using
  {S}emantic {W}eb {S}ervices for business process management.
\newblock In {\em IEEE International Conference on e-Business Engineering
  (ICEBE'05)}, pages 535--540. IEEE, 2005.

\bibitem{hicks2020secureabc}
Chris Hicks, David Butler, Carsten Maple, and Jon Crowcroft.
\newblock Secure{ABC}: {S}ecure {A}nti{B}ody {C}ertificates for {COVID}-19.
\newblock {\em arXiv preprint arXiv:2005.11833}, 2020.

\bibitem{jager2013one}
Tibor Jager, Kenneth~G Paterson, and Juraj Somorovsky.
\newblock One bad apple: Backwards compatibility attacks on state-of-the-art
  cryptography.
\newblock In {\em NDSS}, 2013.

\bibitem{jwt}
M~Jones, J~Bradley, and N~Sakimura.
\newblock {JSON} {W}eb {T}oken ({JWT}).
\newblock {\em IETF RFC 7519}, 2015.

\bibitem{jordan2003augmented}
Ken Jordan, Jan Hauser, and Steven Foster.
\newblock The {A}ugmented {S}ocial {N}etwork: {B}uilding identity and trust
  into the next-generation {I}nternet.
\newblock {\em First Monday}, 8(8), 2003.

\bibitem{kaminer2020discrimination}
Debbie Kaminer.
\newblock Discrimination against employees without {COVID}-19 antibodies.
\newblock {\em New York Law Journal}, 2020.

\bibitem{nature}
Natalie Kofler and Fran{\c{c}}oise Baylis.
\newblock Ten reasons why immunity passports are a bad idea, 2020.

\bibitem{kontzinos2020decentralised}
Christos Kontzinos, Panagiotis Kokkinakos, Stavros Skalidakis, Ourania Markaki,
  Vagelis Karakolis, and John Psarras.
\newblock Decentralised qualifications' verification and management for learner
  empowerment, education reengineering and public sector transformation: {T}he
  {Q}uali{C}hain {P}roject.
\newblock {\em Mobile, Hybrid, and On-line Learning (eLmL 2020)}, page~51,
  2020.

\bibitem{claimchain}
Bogdan Kulynych, Wouter Lueks, Marios Isaakidis, George Danezis, and Carmela
  Troncoso.
\newblock Claimchain: Improving the security and privacy of in-band key
  distribution for messaging.
\newblock In {\em Proceedings of the 2018 Workshop on Privacy in the Electronic
  Society}, pages 86--103, 2018.

\bibitem{larremore2020implications}
Daniel~B Larremore, Kate~M Bubar, and Yonatan~H Grad.
\newblock Implications of test characteristics and population seroprevalence on
  immune passport strategies.
\newblock {\em Clinical Infectious Diseases}, 2020.

\bibitem{w3c:rdf}
Ora Lassila and Ralph~R Swick.
\newblock Resource {D}escription {F}ramework ({RDF}) model and syntax
  specification.
\newblock {\em W3C Recommendation}, 1999.

\bibitem{w3c:linkedproofs}
David Longley and Manu Sporny.
\newblock Linked {D}ata {P}roofs.
\newblock {\em W3C Draft Community Group Report}, 2020.
\newblock \url{https://w3c-ccg.github.io/ld-proofs/}.

\bibitem{mansour2016demonstration}
Essam Mansour, Andrei~Vlad Sambra, Sandro Hawke, Maged Zereba, Sarven
  Capadisli, Abdurrahman Ghanem, Ashraf Aboulnaga, and Tim Berners-Lee.
\newblock A demonstration of the {S}olid platform for social web applications.
\newblock In {\em Proceedings of the 25th International Conference Companion on
  World Wide Web}, pages 223--226. International World Wide Web Conferences
  Steering Committee, 2016.

\bibitem{mcintosh2005xml}
Michael McIntosh and Paula Austel.
\newblock {XML} signature element wrapping attacks and countermeasures.
\newblock In {\em Proceedings of the 2005 workshop on Secure web services},
  pages 20--27. ACM, 2005.

\bibitem{pfitzmann2010terminology}
Andreas Pfitzmann and Marit Hansen.
\newblock A terminology for talking about privacy by data minimization:
  Anonymity, unlinkability, undetectability, unobservability, pseudonymity, and
  identity management.
\newblock 2010.

\bibitem{recordon2006openid}
David Recordon and Drummond Reed.
\newblock Open{ID} 2.0: a platform for user-centric identity management.
\newblock In {\em Proceedings of the second ACM workshop on Digital identity
  management}, pages 11--16. ACM, 2006.

\bibitem{w3c:did}
Drummond Reed, Manu Sporny, and Markus Sabadello.
\newblock {D}ecentralized {I}dentifiers ({DID}s) v1.0.
\newblock {\em W3C Working Draft}, 2020.
\newblock \url{https://www.w3.org/TR/did-core/}.

\bibitem{openidconnect}
N.~Sakimura, J.~Bradley, M.~Jones, B.~de~Medeiros, and C.~Mortimore.
\newblock {OpenID Connect Core 1.0 incorporating errata set 1}, 2014.
\newblock \url{http://openid.net/specs/openid-connect-core-1_0.html}.

\bibitem{w3c:vc}
Manu Sporny, David Longley, and David Chadwick.
\newblock Verifiable {C}redentials.
\newblock {\em W3C Recommendation}, 2019.
\newblock \url{https://www.w3.org/TR/verifiable-claims-data-model/}.

\bibitem{w3c:jsonld}
Manu Sporny, David Longley, Markus Lanthaler, Pierre-Antonine Champin, and
  Niklas Lindstrom.
\newblock {JSON-LD} 1.1: A {JSON} serialization for {L}inked {D}ata.
\newblock {\em W3C Recommendation}, 2020.
\newblock \url{https://www.w3.org/TR/json-ld11/}.

\bibitem{troncoso2017systematizing}
Carmela Troncoso, Marios Isaakidis, George Danezis, and Harry Halpin.
\newblock Systematizing decentralization and privacy: Lessons from 15 years of
  research and deployments.
\newblock {\em Proceedings on Privacy Enhancing Technologies},
  2017(4):404--426, 2017.

\bibitem{wust2018you}
Karl W{\"u}st and Arthur Gervais.
\newblock Do you need a blockchain?
\newblock In {\em 2018 Crypto Valley Conference on Blockchain Technology
  (CVCBT)}, pages 45--54. IEEE, 2018.

\end{thebibliography}

\newpage

\end{document}